\documentclass{ws-p8-50x6-00}

\begin{document}
\newcommand{\pardis}{\langle \mu \rangle}
\newcommand{\diff}{\mbox{d}}
\newcommand{\beq}{\begin{eqnarray}}
\newcommand{\eeq}{\end{eqnarray}}
\newcommand{\proj}{\vec{\Phi}(x)}
\newcommand{\vproj}{\hat{\Phi}(x)}
\newcommand{\latproj}{\hat{\Phi}(\vec{n}+\hat{\imath},t)}
\newcommand{\blatproj}{\hat{\Phi}(\vec{n},t)}
  \newcommand{\epsfaxhax}[2]{
          \centerline{
            \hspace{-15pt}
            \epsfxsize=160pt
            {\epsfbox{#1}}
            \hspace{-7pt}
            \epsfxsize=160pt
            {\epsfbox{#2}}}
  }
\title{What we do understand of colour confinement}
\author{A. Di Giacomo}

\address{Dipartimento di Fisica dell'Universit\`a and INFN, Pisa,
Italy\footnote{e-mail address: digiaco@mailbox.difi.unipi.it}}

\author{B. Lucini}
\address{Scuola Normale Superiore and INFN, Pisa, Italy\footnote{e-mail
address: lucini@cibs.sns.it}} 

\maketitle

\abstracts{The status of our understanding of confinement is reviewed. The
evidence from lattice is that monopole condensation, or dual superconductivity,
is at work. Confinement is an order-disorder transition. Different monopole
species look equivalent, indicating that the symmetry of the disordered phase
is more interesting than we understand.}
\section{Introduction}
No big progress has been made in the last year on the subject.
This is a good time
to assess what we have learned, and to try an outlook. The question in the
title is ambitious. The answer will be: not as much as we think, but we have
good handles.

This paper is the sum of two talks: one presented by the first author on the
general statement of the problem, the other presented by the second author on
specific lattice results. All this is based on results already presented in
ref.'s 1. The topics which will be addressed are:

\hspace{-0.5cm}
a) lattice vs. continuum formulation of QCD;

\hspace{-0.5cm}
b) confinement;

\hspace{-0.5cm}
c) duality and disorder parameter;

\hspace{-0.5cm}
d) monopoles;

\hspace{-0.5cm}
e) monopole condensation, abelian dominance and monopole dominance;

\hspace{-0.5cm}
f) what next.
\section{Lattice vs. continuum}
A popular prejudice is that continuum QCD is based on logical and mathematical
arguments, contrary to lattice, which is based on numerical simulations, and
therefore does not help in understanding. In reality
\begin{enumerate}
\item The continuum quantization is perturbative, has the Fock
vacuum as ground state, and consists in computing scattering processes
between gluons and quarks. Fock vacuum is certainly not the ground state
and this instability is signalled by the presence of renormalons, i.e. by
the fact that the renormalized perturbative expansion is not even an
asymptotic series\cite{Muller}. However, for some reason, which would
be interesting to understand, perturbation theory works at short distances.
\item Lattice formulation is a sensible approximation to the functional
Feynman integral which defines the theory. Most probably QCD exists as a self
consistent field theory, and is defined constructively on a lattice. Gauge
invariance is built in. Since, in addition, objects with non trivial topology
(instantons, monopoles) play an important r\^ole in QCD dynamics, a formulation
in terms of parallel transport, like lattice, is superior.
\item Numerical results are like experiments: what is understood from them
depends on the question they address. Experiments testing a symmetry, like
Michelson-Morley experiment, can be more important to understand than the
computation of 3-loop radiative corrections.
\end{enumerate}
\section{Confinement}
Quarks and gluons have never been observed.
The ratio of quark to nucleons abundance in the universe
is bounded by the experimental limit
\beq
\frac{n_q}{n_p} \le 10^{-27}
\end{eqnarray}
coming from a Millikan like analysis of $\sim 1g$ of matter. In a cosmological
standard model one would expect for the same ratio\cite{Okun}
$n_q/n_p  \simeq 10^{-12}$, which is bigger by 15 orders
of magnitude.

Lattice numerical evidences exist that QCD confines colour. The Wilson
loops obey the area law\cite{Creutz}
\beq
W(R,T) \mathop{\simeq}_{RT \to \infty} \exp( - \sigma RT) \ .
\eeq
Since general arguments imply that
\beq
W(R,T) \mathop{\simeq}_{RT \to \infty} \exp( - V(R)T) \ ,
\eeq
$V(R)$ being the static $Q-\bar{Q}$ potential, it follows that
$V(R) = \sigma R$ ($\sigma$ is the string tension), which means
confinement of quarks.

A guiding principle in our analysis of confinement will be that such an
absolute property like confinement can only be explained in terms of symmetry.
A similar situation exists e.g. in ordinary superconductivity: the resistivity
is observed to be consistent with zero with very great precision.
This is not due to the smallness of a tunable parameter, but to a symmetry.
\section{Duality}
In a finite temperature formulation of QCD the confined phase corresponds to
the strong coupling region (low values of $\beta = \frac{2 N_c}{g^2}$).
Above some $\beta_C$ deconfinement takes place\cite{Deconf}. Apparently the
confined phase is disordered: it should, however, have a nontrivial order, if
confinement has to be explained in terms of symmetry.

A wide class of systems exist in statistical mechanics and in field theory,
in which a similar situation occurs\cite{Duali,Nonduali}.
All those systems have topologically non trivial configurations, carrying a
conserved topological charge, and admit two equivalent descriptions (duality).
The ``ordered'' phase (low values of $\beta$)
is described in terms of the usual local fields, and its symmetry is
discussed in terms of vacuum expectation values ({\em vev}) of local
fields (order parameters): in this description topological excitations
are extended objects (kinks, vortices, $\dots$). In the dual description
the extended topological objects are described in terms of dual local fields,
the coupling constant is $\sim 1/g$ and the disordered phase looks
ordered, while the phase that originally was ordered looks disordered.
The order parameter of the dual description is called a disorder parameter.
Understanding confinement consists then in understanding the symmetry of the
system dual to QCD. In fact the explicit construction of the dual system is
not required: once the dual symmetry is understood the disorder parameter can
be constructed in terms of the original local fields, of course as a highly
non local operator\cite{Nonduali}.

A suggestive possibility in this direction is that QCD vacuum could behave as
a dual superconductor\cite{Mandelstam}. Confinement would be a consequence of
the dual Meissner effect, squeezing the chromoelectric field of a $Q-\bar{Q}$
pair into an Abrikosov flux tube with energy proportional to the length,
or
\beq
V(R) = \sigma R \ .
\eeq
In this mechanism monopoles are the topological structures which are expected
to condense.

Chromoelectric flux tubes are indeed observed on lattice
configurations\cite{Haymaker} and also their collective modes have been
detected\cite{Collettivi}. The main results of a more detailed
analysis of this mechanism will be presented below.
\section{Monopoles in non abelian gauge theories}
We shall refer to $SU(2)$ for simplicity: the
extension to $SU(3)$ only involves some additional formal complications.

Let $\Phi = \vec{\Phi} \cdot \vec{\sigma}$ be any operator in the adjoint
representation. A unit colour vector $\vproj$ can be defined:
\beq
\vproj = \frac{\proj}{\mid \proj \mid}
\eeq
everywhere except at sites where $\proj = 0$. The field configuration
$\proj$ can present a non trivial topology. If we adopt a ``local''
reference frame for colour, with 3 orthonormal unit vectors $\vec{\xi}_i(x)$,
$\vec{\xi}_i(x) \cdot \vec{\xi}_j(x) = \delta_{ij}$, $\vec{\xi}_i(x) \wedge
\vec{\xi}_j(x) = \epsilon _{ijk} \vec{\xi}_k(x)$, with $\vec{\xi}_3(x) = 
\vproj$, instead of the usual $x$ independent unit vectors $\vec{\xi}_{i}^o$,
a rotation $R(x)$ will exist such that
\beq
\label{eq:6}
\vec{\xi}_i(x) = R(x) \vec{\xi}_i^o \ .
\eeq
Since $\mid \vec{\xi}_i(x) \mid ^2 = 1$,
\beq
\partial _{\mu} \vec{\xi}_i(x) = \vec{\omega}_{\mu} \wedge \vec{\xi}_i(x)
\eeq
or
\beq
\label{eq:8}
D_{\mu} \vec{\xi}_i(x) = \left( \partial _{\mu} - \vec{\omega}_{\mu} \wedge
\right) \vec{\xi}_i(x) = 0 \ .
\eeq
The symbol $\wedge$ indicates cross product; the $SO(3)$ generators are in the
fundamental representation $T_{ij}^a = - i \epsilon _{iaj}$.

Eq. (\ref{eq:8}) implies $[D_{\mu},D_{\nu}] = 0$, or
\beq
\label{eq:9}
\vec{F}_{\mu \nu} = \partial _{\mu} \vec{\omega}_{\nu} - \partial _{\nu}
\vec{\omega}_{\mu} + \vec{\omega}_{\mu} \wedge \vec{\omega}_{\nu} = 0 \ .
\eeq 
The rotation (\ref{eq:6}) is a parallel transport, and as such it is a pure
gauge.

The solution of eq.~(\ref{eq:8}) is
\beq
\label{eq:10}
\vec{\xi}_i(x) = \mbox{Pexp} \left( i \int _{\infty, C}^x \vec{\omega}_{\mu}
\cdot \vec{T} \,d x^{\mu} \right) \vec{\xi}_i^o \ ,
\eeq
and eq.~(\ref{eq:9}) implies that the path integral~(\ref{eq:10}) is
independent of the choice of the line $C$.
In fact eq. (\ref{eq:9}) is not valid at the singularities
occurring at the zeros of $\proj$, where $R(x)$ is not defined, and
as a consequence $\vec{\xi}_i(x)$ is not independent of the path $C$.

The inverse rotation $R^{-1}(x)$ acts on $\vproj =  \vec{\xi}_3$ as
\beq
R^{-1}(x) \vproj =  \vec{\xi}_3^o \ .
\eeq 
$R^{-1}(x)$ is called abelian projection.

Under abelian projection the field strength tensor $\vec{G}_{\mu \nu} = 
\partial_{\mu} A_{\nu} - {\partial}_{\nu} A_{\mu} + g A_{\mu} \wedge A_{\nu}$
has the usual covariant transformation out of the singularities. Where
$R(x)$ is not defined it can acquire a singular term
\beq
\label{eq:15}
\vec{G}_{\mu \nu} \mathop{\to}_{R^{-1}(x)} 
\vec{G}_{\mu \nu} + \vproj \left( \partial _{\mu}
A_{\nu} ^{sing} - \partial_{\nu} A_{\mu} ^{sing} \right) \ .
\eeq 

The quantity\cite{'tHooft1}
\beq
F_{\mu \nu} = \vproj \cdot \vec{G}_{\mu \nu} - \frac{1}{g}
\left( D_{\mu} \vproj \wedge D_{\nu} \vproj \right) \cdot \vproj 
\eeq
can be identically put in the form
\beq
F_{\mu \nu} =  {\partial}_{\mu} A_{\nu} - {\partial}_{\nu} A_{\mu} -\frac{1}{g}
\left( \partial _{\mu} \vproj \wedge \partial _{\nu} \vproj 
\right) \cdot \vproj \ .
\eeq
In the abelian projected form the second term disappears, since $\vproj \to
\vec{\xi}_3^o$ which is $x$ independent and
\beq
F_{\mu \nu} =  {\partial}_{\mu} A_{\nu} - {\partial}_{\nu} A_{\mu} + \mbox{
singular term} \ .
\eeq
$F_{\mu \nu}$ is an abelian field. 

$F^{\star}_{\mu \nu} = \frac{1}{2} \epsilon _{\mu \nu \rho \sigma} F^{\rho 
\sigma}$ defines a magnetic current
\beq
j_{\mu}^M &=& {\partial}^{\nu} {F}^{\star}_{\mu \nu} \ ,
\eeq
which is identically conserved. The system has a magnetic $U(1)$ symmetry.
If there are no singularities $j_{\mu}^M$ itself vanishes (Bianchi identities).
It can be shown that the singularities of the abelian projection
are nothing but pointlike $U(1)$ magnetic charges: the singular term in eq.
(\ref{eq:15}) is a Dirac string taking care of flux conservation.

A magnetic $U(1)$ symmetry exists for each of the functionally-infinite
choices of $\proj$: one can construct a disorder parameter as
the {\em vev} of an operator carrying non zero magnetic charge, and
investigate dual superconductivity. The disorder parameter can be then
measured on the lattice, and the symmetry of the confined vacuum can be
determined. This has been done for a number of different choices of the
operator $\proj$, and for all of them the transition to confined
phase is a transition from dual normal conductor to dual superconductor.
\section{Monopole condensation: the disorder parameter}
The disorder parameter is the {\em vev} of a non local operator.
The logical procedure to define this operator merely consists in
shifting the field variables at a given time $t$ by a configuration
describing a static monopole sitting at $\vec{y}$, $\vec{A}(\vec{x},\vec{y})$,
in the same way as for a particle moving in one dimension
\beq
e^{i p a} | x \rangle = | x + a \rangle \ . 
\eeq 
For a field configuration $\vec{A}(\vec{x},t)$, adding the field
$\vec{A}(\vec{x},\vec{y})$ amounts to
\beq
\mu | \vec{A}(\vec{x},t) \rangle =
e^{i \int \vec{\pi}(\vec{x},t) \cdot \vec
{A}(\vec{x},\vec{y}) \,d \vec{x}} | \vec{A}(\vec{x},t) \rangle
= | \vec{A}(\vec{x},t) + \vec{A}(\vec{x},\vec{y}) \rangle \ .
\eeq
In doing that care has to be taken of the compactness of the theory, and of
the fact that we want to add a monopole to the abelian part of the abelian
projected field. All this can be done exactly and gives\cite{Giappone}
\beq
\label{eq:19}
\pardis = \frac{Z[S+\Delta S]}{Z [ S ] } \ ,
\eeq
where $Z$ is the usual partition function of the theory and $\Delta S$
consists in a modification to the action $S$ in the slice $x_0 = t$, in all
points of space (non local operator).

The recipe is to change the temporal plaquette $\Pi_{i0}(\vec{n},t)$
to $\Pi_{i0}^{\prime}(\vec{n},t)$:
\beq
\Pi _{i 0} (\vec{n},t) = U_i(\vec{n},t)U_0(\vec{n}+ \hat \imath,t)
\left(U_i (\vec{n},t+1)\right)^{\dag}\left(U_0(\vec{n},t)\right)^{\dag} 
\eeq
\beq
\label{eq:3.15}
\Pi _{i 0}  ^{\prime} (\vec{n},t)
= U_i^{\prime}(\vec{n},t)U_0(\vec{n}+ \hat \imath,t)
\left(U_i (\vec{n},t+1)\right)^{\dag}
\left(U_0(\vec{n},t)\right)^{\dag} \ ,
\eeq
\beq
\label{eq:3.16}
U_i^{\prime}(\vec{n},t) &=& e^{- i \Lambda (\vec{n},\vec{y}) \blatproj \cdot
\vec{\sigma}}
U_i(\vec{n},t) e^{i A_{\bot i}^M (\vec{n}+\hat{\imath}/2,\vec{y})
\latproj \cdot \vec{\sigma}}\\
&&e^{i \Lambda (\vec{n}+\hat{\imath},\vec{y}) \latproj \cdot
\vec{\sigma}}  \ ,
\eeq
$ \vec{A}_{\bot}^M(\vec{x},\vec{y})$ and $\Lambda(\vec{x},\vec{y})$ being
respectively the transverse ($\vec{\nabla} \cdot \vec{A}_{\bot}^M(\vec{x},
\vec{y}) = 0$) and the pure gauge part ($\vec{\nabla} \Lambda(\vec{x},\vec{y})=
A_{\parallel}(\vec{x},\vec{y})$) of $\vec{A}(\vec{x},\vec{y})$.

The operator is in fact $\mu = e^ {- \beta \Delta S}$, with
$\Delta S \sim N_s^3$, $N_s$ being the spatial extension of the system. The
fluctuations of $\pardis$ are then $\sim \exp(N_s^{3/2})$. Instead of
$\pardis$, which is a widely fluctuating quantity, it proves to be convenient
to define
\beq
\rho = \frac{\diff}{\diff \beta} \log \pardis \ .
\eeq

Eq. (\ref{eq:19}) gives
\beq
\label{rhoeq}
\rho = \langle S \rangle _S - \langle S + \Delta S \rangle _{S + \Delta S} \ .
\eeq
The subscript denotes the action used in weighting the average.
\begin{figure}[t]
\epsfaxhax{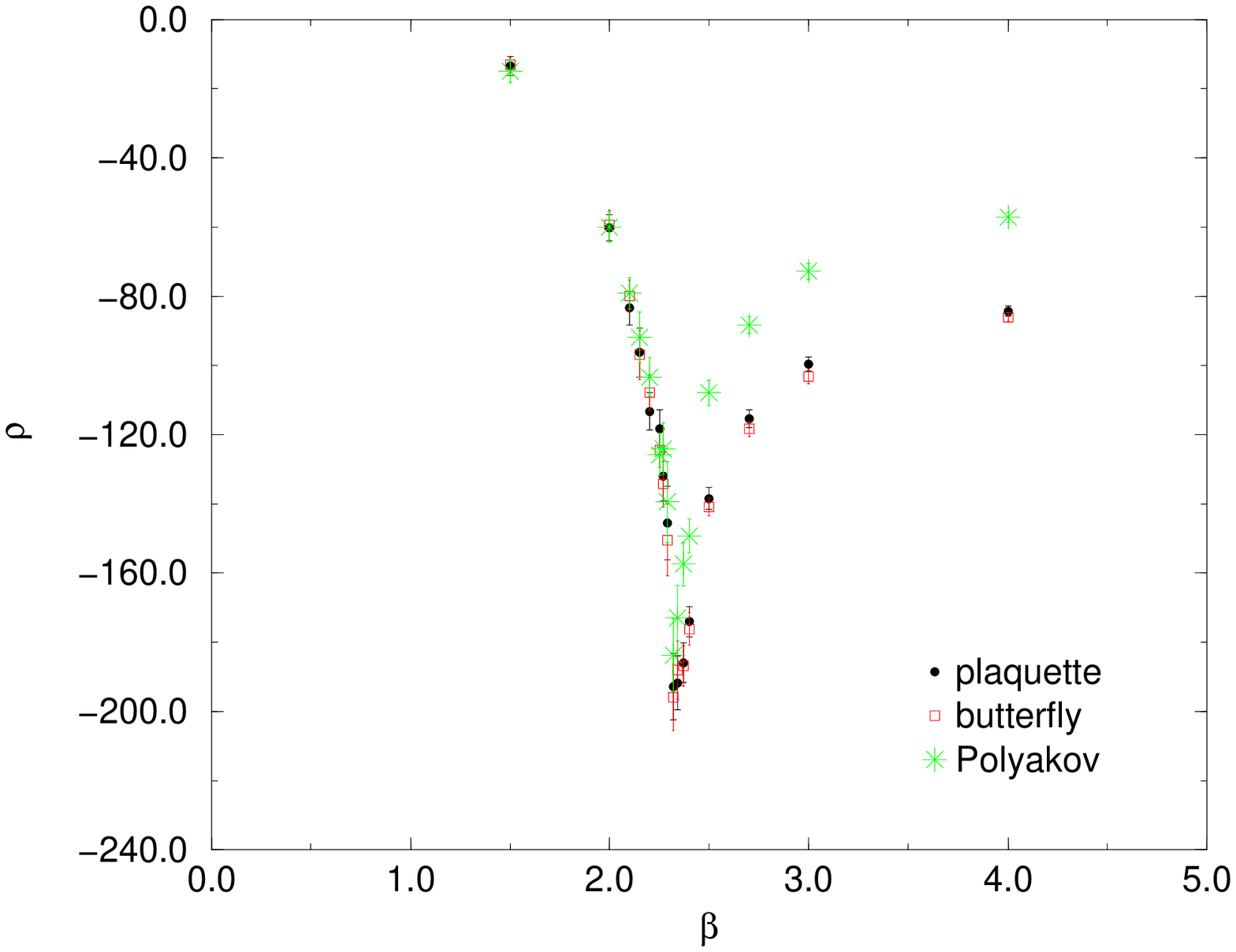}{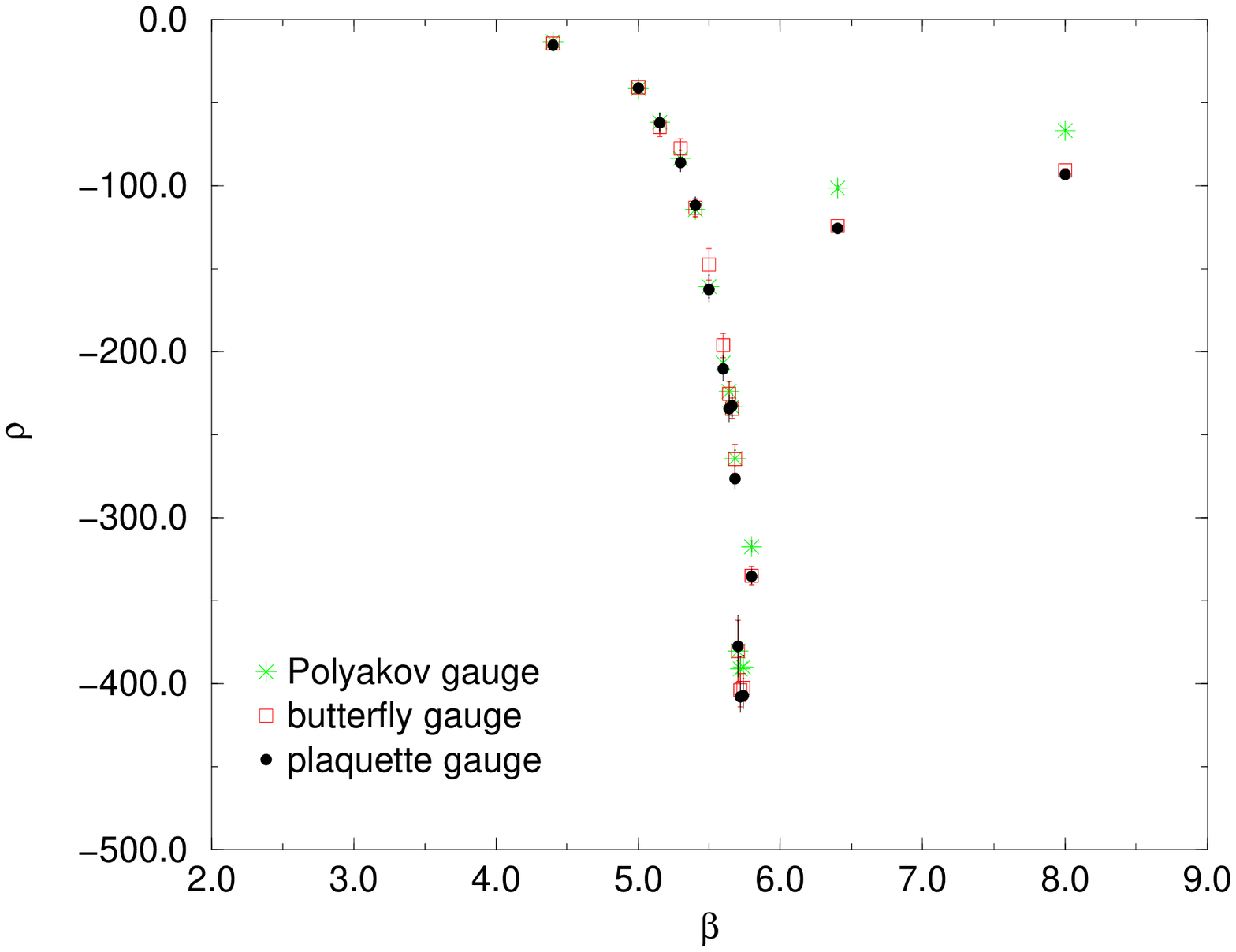}
\vskip -0.3cm
\begin{center}
\mbox{\tiny{(a)}}\mbox{\hskip 5.3cm}\mbox{\tiny{(b)}}
\end{center}
\caption{$\rho$ vs. $\beta$ in different abelian projections for (a) $SU(2)$ 
and (b) $SU(3)$.}
\label{fig1}
\vskip -0.3cm
\setcounter{figure}{1}
\end{figure} 
\begin{figure}[h]
\epsfaxhax{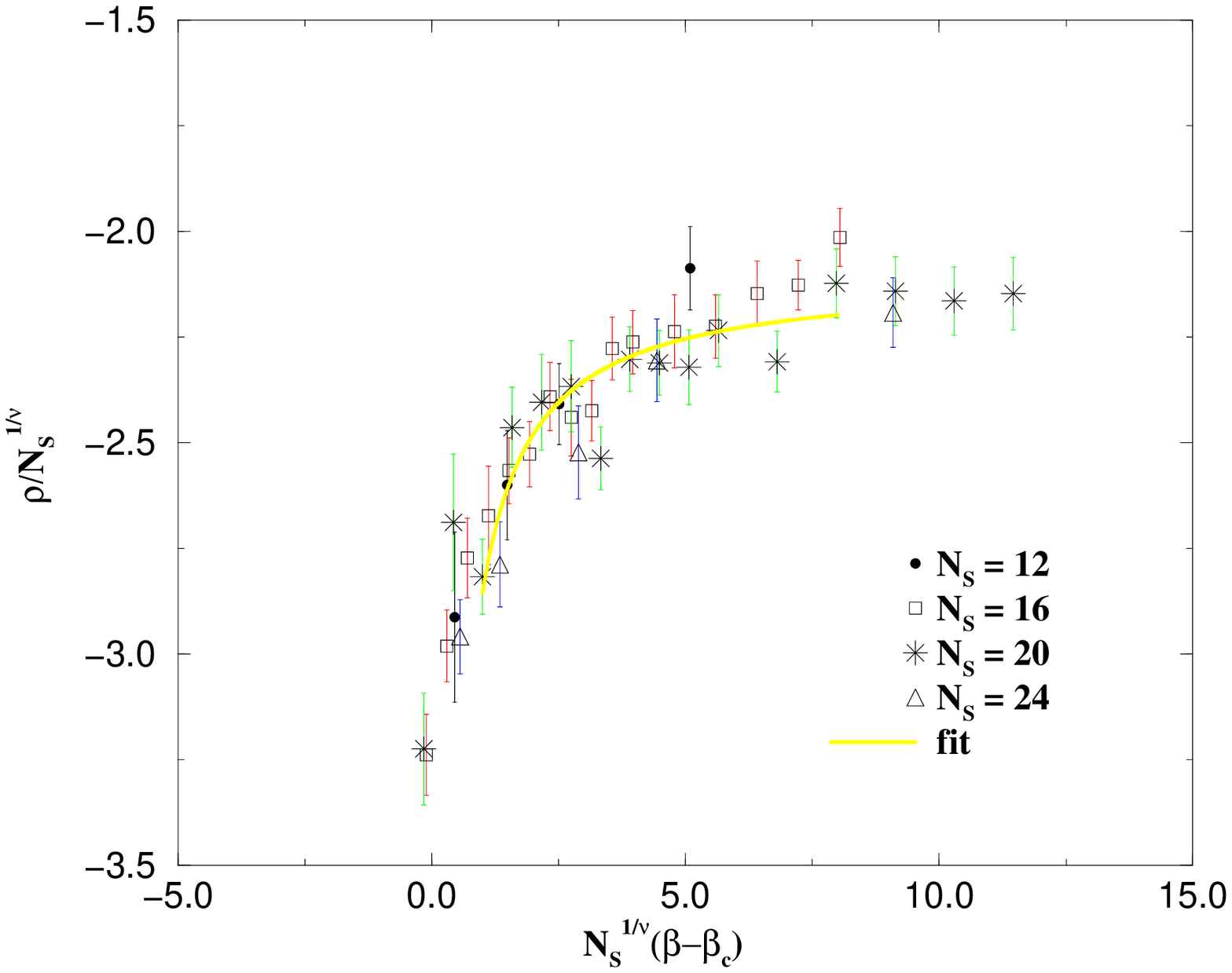}{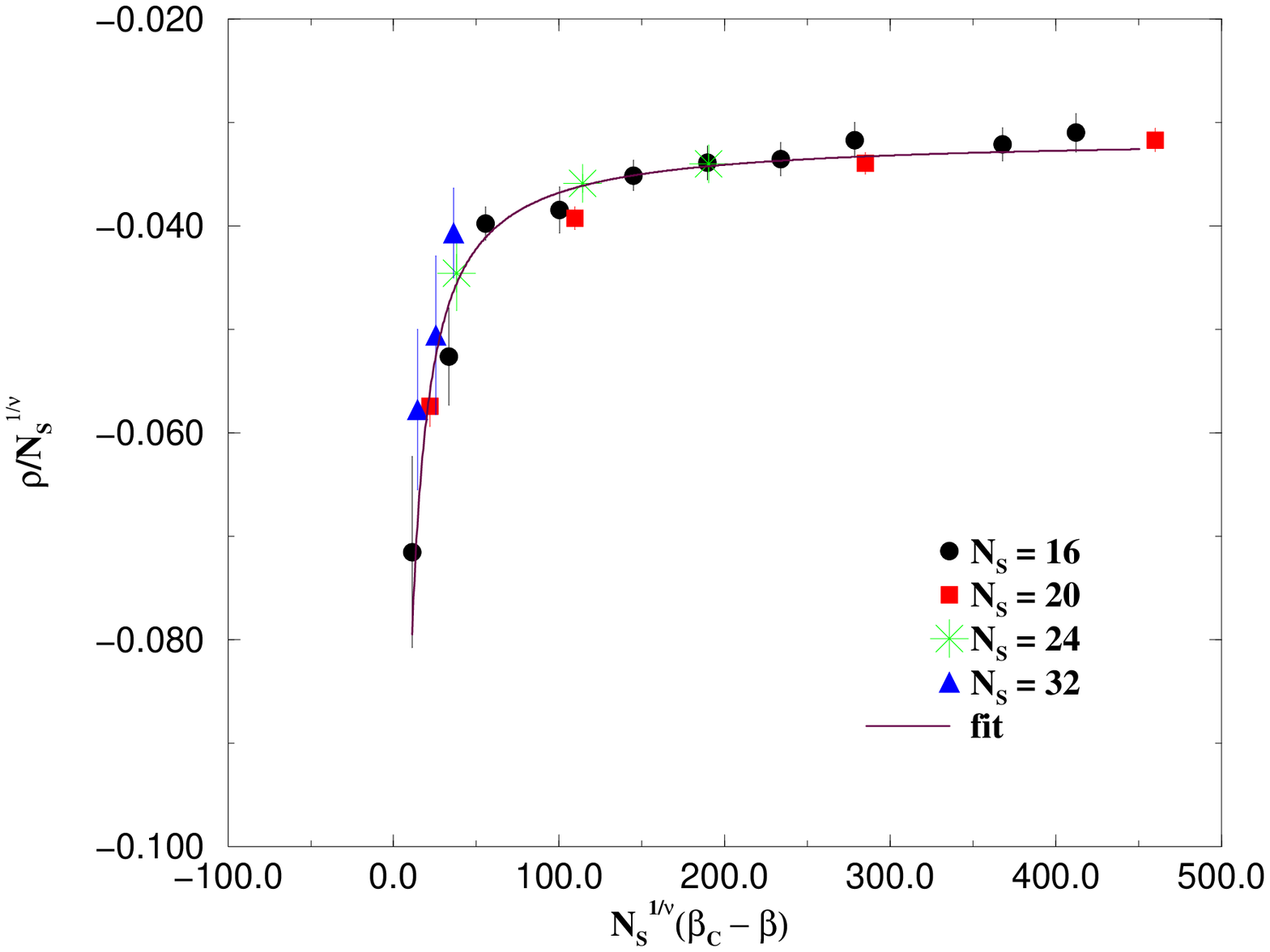}
\vskip -0.3cm
\begin{center}
\mbox{\hskip0.3cm}\mbox{\tiny{(a)}}\mbox{\hskip 5.3cm}\mbox{\tiny{(b)}}
\end{center}
\label{fig4}
\caption{Rescaled $\rho$ data as a function of the scaling variable
for (a) $SU(2)$ and (b) $SU(3)$.}
\vskip -0.3cm
\setcounter{figure}{4}
\vskip -0.3cm
\end{figure} 
\section{Monopole condensation in $SU(2)$ and $SU(3)$}
We studied numerically on the lattice the deconfining phase transition
at finite temperature for $SU(2)$ and $SU(3)$ by means of the quantity $\rho$,
eq. (\ref{rhoeq}). Finite temperature means that our lattice is $N_s^3 \times
N_t$, with $N_s \gg N_t$. $N_s^3$ is the physical volume, while $N_t$ is
related to the temperature $T$ by the relationship $T = 1/\left[N_t
a(\beta)\right]$.
At finite temperature, $C^{\star}$ boundary conditions\cite{Wiese} have
to be used in the $t$ direction.

We investigated condensation for the monopoles defined by the following
operators $\Phi$:
\begin{itemize}
\item $\Phi$~is~related~to~the~Polyakov~line~$L(\vec{n},t) =
\Pi _{t^{\prime}=t}^{N_t -1} U_0(\vec{n},t^{\prime})
\Pi _{t^{\prime}=0}^{t-1} U_0(\vec{n},t^{\prime})$ in the following way:
\beq
\label{eq:4.29}
\Phi(n) \equiv \Phi(\vec{n},t) = L (\vec{n},t) L^{\star} (\vec{n},t)  
\eeq
(Polyakov projection on a $C^{\star}$ periodic lattice\footnote{the symbol
$\star$ in eq. (\ref{eq:4.29}) indicates the complex conjugation operation.});
\item $\Phi$ is an open plaquette, i.e. a parallel transport on an elementary
square of the lattice
\beq
\label{eq:4.30}
\Phi(n)= \Pi _{ij}(\vec{n},t) = U_i(n)U_j(n+\hat{\imath})
\left(U_i(n+\hat{\jmath})\right) ^{\dag} \left(U_j(n) \right) ^{\dag}\ ;
\eeq
\item $\Phi$ is the ``butterfly'' (topological charge density) operator:
\beq
\label{eq:4.31}
\Phi(n) =  F (\vec{n},t) = U_x(n)U_y(n+\hat{x})
\left(U_x(n+\hat{y})\right)^{\dag} \left( U_y(n)\right)^{\dag}\\
\nonumber
U_z(n)U_t(n+\hat{z}) \left(U_z(n+\hat{t})\right)^{\dag}
\left(U_t(n)\right) ^{\dag} \ .
\eeq
\end{itemize}

Our main results are
\begin{enumerate}
\item $\rho$ shows a sharp negative peak in the critical region
independently of the abelian projection chosen (fig.~\ref{fig1}).
$\pardis$ has an abrupt decline in the critical
region. $\rho$ does not depend on the abelian projection used.
\item The peak position follows the displacement of
the critical temperature when the size in the $t$ direction is changed.
\item In the $SU(3)$ case for a given abelian projection we have two possible
choices for defining monopoles, the residual gauge group being $[U(1)]^2$.
The corresponding $\rho$'s show a similar behaviour.
\item For both $SU(2)$ and $SU(3)$ at strong coupling
(low $\beta$'s) $\rho$ seems to reach a finite limit when $N_s \to \infty$.
From the condition
\beq
\label{eq:4.33}
\pardis  = \exp \left( \int _0^{\beta} \rho (\beta^{\prime}) \diff
\beta ^{\prime} \right) 
\eeq
it follows that $\pardis \ne 0$ and the magnetic $[U(1)]^{N-1}$ symmetry is
broken in the confined phase.
\item At weak coupling (large $\beta$'s) numerical data are consistent
with a linear behaviour of $\rho$ as a function of $N_s$.
For $\pardis$ we get
\beq
\langle \mu \rangle \mathop{\approx}_{N_s \to \infty} 
A e^{ (- c N_s + d )\beta } \ ,
\eeq 
with $c \simeq 0.6$ and $d \simeq -12$ for $SU(2)$ and $c \simeq 2$ and $d
\simeq -12$ for $SU(3)$.
$\pardis = 0$ in the deconfined phase.
\item A finite size scaling analysis shows that the disorder parameter
reproduces the correct critical indices and the correct $\beta_C$
in both cases. We have indeed
\beq
\label{scaling} 
\frac{\rho}{N_s^{1/\nu}} = f\left( N_s^{1/\nu} \left( \beta _C - \beta
\right)\right) \ ,
\eeq
i.e. $\rho/N_s^{1/\nu}$ is a function of the scaling variable
$x = \left( N_s^{1/\nu} \left( \beta _C - \beta \right) \right)$.
In eq. (\ref{scaling}) $\nu$ is the critical index associated to the
divergence of the correlation length (pseudo-divergence for a first
order phase transition) and $\beta_C$ is the critical value of $\beta$.
We parameterize $\rho/N_s^{1/\nu}$ as
\beq
\frac{\rho}{N_s^{1/\nu}} = - \frac{\delta}{x}  - c  + \frac{a}{N_s^3} \ ,
\eeq
where $\delta$ is the exponent associated to the drop of $\pardis$
at the transition ($\pardis \mathop{\propto} \left( \beta _C - \beta \right)
^{\delta}$, ${N_s \to \infty}$),
$c$ is a constant and $a$ measures scaling violations. We find
independently of the abelian projection and (for $SU(3)$) of the abelian
generator $\beta_C = 2.29(3)$, $\nu = 0.63(5)$, $\delta = 0.20(8)$, $a \sim 0$
for $SU(2)$ and $\beta_C = 5.69(3)$, $\nu = 0.33(5)$,
$\delta = 0.54(4)$, $a \sim 210$ for $SU(3)$ (fig.~2).
$\beta_C$ and $\nu$ agree with ref.'s 14.
\end{enumerate}
\section{Discussion}
The firm point we have is that confinement
is an order disorder transition. Whatever the dual symmetry of the disordered
phase is, the operators defining dual order have non
zero magnetic charge in different abelian projections.

For sure the statement that only one abelian projection is at work, e.g.
the maximal abelian, is inconsistent with this symmetry. May be maximal abelian
is more convenient than other abelian projections to build effective
lagrangeans, but this point is not directly relevant to symmetry.

In addition, if one single abelian projection were involved in monopole
condensation, then flux tubes observed on lattice in $Q-\bar{Q}$ configurations
should have the electric field belonging to the confining $U(1)$, and this is
not the case\cite{tubi}. Moreover there exist coloured states which are neutral
with respect to that $U(1)$, and they would not be confined.

Symmetry of the dual field theory is more clever than we think. It is any how
related to condensation of monopoles in different abelian projections. As
guessed in ref. 16 all abelian projections are physically
equivalent.

\end{document}